\documentclass[journal]{IEEEtran}

\usepackage{amsmath}
\usepackage{amssymb}
\usepackage{graphicx}
\usepackage[boxed]{algorithm2e}
\usepackage{color}
\begin{document}
\title{Optimization of Non Binary Parity Check Coefficients}

\author{\IEEEauthorblockN{Emmanuel Boutillon,~\IEEEmembership{Senior~Member,~IEEE} }\\ 
\IEEEauthorblockA{Lab-STICC, UMR 6582, Universit\'e de Bretagne Sud 56100 Lorient, France\\ email: emmanuel.boutillon@univ-ubs.fr}} %

\maketitle

\begin{abstract}
This paper generalizes the method proposed by Poulliat et al. for the determination of the optimal Galois Field coefficients of a Non-Binary LDPC parity check constraint based on the binary image of the code. Optimal, or almost-optimal, parity check coefficients are given for check degree varying from 4 to 20 and Galois Field varying from GF(64) up to GF(1024). For all given sets of coefficients, no codeword of Hamming weight two exists. A reduced complexity algorithm to compute the binary Hamming weight 3 of a parity check is proposed. When the number of sets of coefficients is too high for an exhaustive search and evaluation, a local greedy search is performed. Explicit tables of coefficients are given. The proposed sets of coefficients can effectively replace the random selection of coefficients often used in NB-LDPC construction. 
\end{abstract}

.

\section{Introduction}
\label{sec:intro}

Non-Binary Low Density Parity Check Codes (NB-LDPC) have been proposed by Mackay and Neal in 1996 as a generalization of the LDPC matrices \cite{Mackay1996}. In \cite{Poulliat2008}, Poulliat et al. present in 2008 a method to set the non-zero coefficients of a non binary parity check matrix $H$. The first step of the method concerns the problem of row optimization, i.e, the selection of the coefficients associated to a given parity check. The principle is to optimize the Hamming weight spectrum of the binary code $(md_c, m(d_c - 1))$ associated to a parity check of degree $d_c$ over a Galois Field GF($q$) with $m = \log_2(q)$. The authors show that the higher the minimum distance of the binary equivalent code, the better is the convergence of the NB-LDPC code in the waterfall region. They also show that, for two parity checks with the same associated binary minimum distance $d_H$, the multiplicity of binary codewords of Hamming distance $d_H$ verifying the parity check equation should be minimized. Once the coefficients of the parity check equation are selected, the second step of \cite{Poulliat2008} is to enumerate the cycles of short lengths in the Tanner graph associated to the parity check matrix and to constraint the GF($q$) coefficients associated to each cycle so that only the zero codeword is associated to the short length cycles. This second step is out of the scope of this paper. The state of the art on coefficients selection is quite sparse, except in \cite{website2003} and \cite{Dolecek2014}. In \cite{website2003}, Mackay proposed to select the set of non null coefficients that maximizes the marginal entropy of one element of the syndrome vector. In \cite{Dolecek2014}, a method used to construct the NB-LDPC code used by the Consultative Committee for Space Data Systems (CCSDS) are presented and some sets of coefficients for $d_c = 4$ over GF(64) and GF(256) are given. We should also mention the paper of \cite{Brouwer1993} which shows minimum Hamming distance upper bound of short length binary codes.  

A direct exploration of all possible codewords associated to a given set of coefficients is limited to small check node degree and Galois Field order due to the exponential increase of complexity. In fact, the number of codewords for a parity check of degree $d_c$ over GF($q$) is $q^{d_c-1}$. For example, for $d_c = 5$ over GF(64), there is $64^4 = 16.8\times 10^6$ codewords per set of coefficients. The number of sets of coefficients is around $8 \times 10^4$ (see section \ref{sec:complexity}): the direct method shows rapidly its limit since it requires more that 100 billions of operations. In \cite{Poulliat2008}, optimal, or almost optimal, sets of coefficients are only given for $d_c = 4$ over GF(64), GF(128) and GF(256).

In this paper, we revisit the problem of coefficient optimization in the case where the binary hamming weight associated to the parity check is strictly greater than 2. We propose a method with a complexity of $\mathcal{O}(d_c^2)$ to evaluate the number of codewords of weight 3. When the number of sets of coefficients is too high for an exhaustive search, a local greedy search is performed. Explicit tables of coefficients are given for $d_c$ varying from 3 to 20 and for Galois Field GF(64) up to GF(1024). The proposed sets of coefficients can effectively replace the random selection of coefficients often used in NB-LDPC construction. For example, let us consider a check node of degree 12 over GF(256), then, in average, randomly selected set of coefficients leads to 68 codewords of weight 3 while the optimized set of coefficients has only 11 codewords of weight 3. In other words, using proposed coefficients, each parity check equation has a better individual error correction, leading globally to a better convergence in the waterfall of the whole NB-LDPC code. 

The remainder of the paper is organized as follows. Section II presents the background on the parity check equation. Section III  states the optimization problem and proposes an effective method to find optimal, or optimized, sets of coefficients. Finally, section IV concludes the paper. All the sets of optimal/optimized coefficients are given in the Appendix II.

\section{Fundamental properties of NB-Parity Check spectrum} \label{sec:soa}

The Galois Field GF($2^m$) will be represented by the set of polynomials over GF(2) modulo $P_m[X]$, where $P_m[X]$ is an irreducible polynomial of degree $m$. Thus, by definition, GF($2^m$) = GF($2$)$[X]/P_m[X]$. It is usual to represent an element of this field either by setting $X = \alpha$ and representing the non null element as power of $\alpha$, i.e, if $x \in$ GF($q$), then $x \neq 0$ implies that $x$ can be written as $x = \alpha^a$, with $a$ a natural that takes its value between 0 and $q-2$.  It is also possible to represent an element of GF($q$) by a binary vector of size $m$ that represents the coefficients of 
a polynomial of GF($2$)$[X]/P_m[X]$ over the base $(1, \alpha, \ldots \alpha^{m-1})$. In this paper, we use the following irreducible polynomials to construct the Galois Field of size 64 up to 1024. 

\begin{equation}
\left\{
 \begin{array}{rll}
P_{6}[X] &= 1 + X + X^6 \\
P_{7}[X] &= 1 + X^3 + X^7 \\
P_{8}[X] &= 1 + X^2 + X^3 + X^4 + X^8 \\
P_{9}[X] &= 1 + X^5 + X^9 \\
P_{10}[X] &= 1 + X^4 + X^{10} \\ 
\end{array}
\right.
\end{equation}

A parity check code $\mathcal{C}$ of degree $d_c$ over GF($q$)$^{d_c}$ is a code defined by a set of $d_c$ non-null GF($q$) coefficients $H = \{ h_i\}_{i=1,2, \ldots d_c}$, with $h_i = \alpha^{a_i}$. Vector $\bold{X} = (x_1, x_2, \ldots, x_{d_c})$ of GF($q$)$^{d_c}$ belongs to the code $\mathcal{C}$ if and only if

\begin{equation}
   h_1x_1 + h_2x_2 + \ldots + h_{d_c}x_{d_c} = 0,
	\label{eq:pc}
\end{equation}
	
\noindent where additions and multiplications are done in GF($q$). Since addition in GF($q$) is commutative, the order of the coefficients does not impact the properties of the code \cite{Poulliat2008}. Moreover, multiplying (\ref{eq:pc}) by a constant factor does not change the equation \cite{Poulliat2008}. In other words, we can always select the coefficients of a parity check code $\mathcal{C}$ so that $h_i = \alpha^{a_i}$ verifies $h_1 = \alpha^0$ (or $a_1 = 0$) and $i \leq j \Rightarrow a_i \leq a_j$. In the sequel, this convention will be used by default.  

Since $\bold{X} \in \mathcal{C}$ is a vector of $GF($q$)^{d_c}$, it is possible to determine its binary image to define a binary code of length $(md_c, m(d_c - 1))$. The Hamming weight spectrum $\mathcal{S}[X]$ of this code is defined as

\begin{equation}
   \mathcal{S}[X] = 1 + S_1X + S_2X^2 + S_3X^3 + \ldots + S_{md_c}X^{md_c},
	\label{eq:hammming}
\end{equation}

\noindent where $S_n$ is the total number of codewords of Hamming weight $n$ of the code. By convention, for a given set of coefficients $H$, $S_n(H)$ will denote the value of the $n^{th}$ coefficient of the Hamming weight spectrum of the code defined by the set of coefficients $H$. The computation of the spectrum can be performed with a complexity of $q^2(d_c-1) + 2q$ using the recursive algorithm used to compute the spectrum distance of a convolutional code \cite{Costello1989}. The adaptation of the algorithm is given in Algo. \ref{al:calcul_spectrum}. The partial spectrum $\mathcal{S}_y(l)[X]$, with $y \in$ GF($q$), $l = 0, 1, 2, \ldots d_c$ represents the spectrum of codewords $(x_1, x_2, \ldots x_l)$ of size $l$ that verify 

\begin{equation}
   \sum_{i=1}^l h_ix_i = y.
	\label{eq:pc_partiel}
\end{equation}

\noindent Note that when $l = 0$, we will assume that $\mathcal{S}_y(0)[X] = 1$ if $y = 0$ (empty set is a solution), 0 otherwise (there is no solution).

Moreover, it is possible to associate also a Hamming Spectrum $\mathcal{S}^x[X]$ to an element of $x \in$ GF($q$). It is the monomial $\mathcal{S}^x[X] = X^{W(x)}$ where $W(x)$ is the binary Hamming weight of $x$, i.e., the number of 1 in the polynomial representation of $x$.

\begin{algorithm}
 {
 \KwData{Initial set of coefficients $H$}
 \KwResult{Spectrum $\mathcal{S}[X]$}
 $\mathcal{S}_s(0)[X] = 1$ if $s = 0$, 0 otherwise. \\
 \For{$l=1, \ldots d_c$}
	 {
	\For{$d \in$ GF($q$)}   
				 { 
				  $\mathcal{S}_d(l)[X] = 0$
				 }
	  
		\For{$s \in$ GF($q$)}   
				 {
  				\For{$x \in$ GF($q$)}  
					    {
							 $d = s + (h_lx)$; \\
               $\mathcal{S}_d(l)[X]$ += $\mathcal{S}_s(l-1)[X] \mathcal{S}^x[X]$;
							}
				 }
		}
		$\mathcal{S}[X]  = \mathcal{S}_0(d_c)[X]$ 
 }
\caption{Computation of Hamming weight spectrum associated to a parity check code}
\label{al:calcul_spectrum}
\end{algorithm}

In an Additive White Gaussian Noise (AWGN) channel, it is well known that the performance is determined first by the Hamming distance of the code (the index $d_{min}$ of the smallest non null value of $\mathcal{S}[X]$, i.e. $S_{d_{min}} \neq 0$ while $0< i <d_{min} \Rightarrow S_i = 0$) and second by the multiplicity of code word of minimum Hamming distance, i.e., the value $S_{d_{min}}$.\\

\noindent \textbf{Lemma 1:} Let $H$ be a set of $d_c$ non-null coefficients in GF($2^m$)$^{d_c}$, then

\begin{equation}
		S_1(H) = 0.
  	\label{eq:s1}
\end{equation}
\\

\noindent \textbf{Lemma 2:} Let $H$ be a set of $d_c$ non-null coefficients in GF($2^m$)$^{d_c}$, then

\begin{equation}
		S_2(H) = \sum_{i=1}^{d_c - 1}\sum_{j=i+1}^{d_c} S_2(\{h_i, h_j\}),
		\label{eq:s2}
		\end{equation}				
\\
\noindent where $S_2(\{h_i, h_j\})$ denotes the number of codewords $(x_i, x_j)$ satisfying the reduced parity check equation $x_ih_i + x_jh_j = 0$ of Hamming weight two.

\noindent  \textbf{Proof:} If $(x_i, x_j)$ is a Hamming weight two solution of $x_ih_i + x_jh_j = 0$, then a vector $X$ having 0 value in all positions, except $X(i) = x_i$ and $X(j) = x_j$ is also a Hamming weight two solution of (\ref{eq:pc}). The total number of codewords of binary Hamming weight two is thus the summation of the number of codewords of binary Hamming weight two associated to each distinct couple of coefficients $\blacksquare$  \\ 

\noindent \textbf{Lemma 3:} Let $H$ be a set of $d_c$ non-null coefficients in GF($2^m$)$^{d_c}$, then

\begin{align}
	 S_3(H) =  S^t_3(H) - (d_c - 3)S^c_3(H) 
		\label{eq:s3}
\end{align}
\noindent where the term $S^t_3(H)$ indicates the summation of binary Hamming weight 3 associated to all possible triplets of non-null coefficients, i.e.,  

\begin{align}
	S^t_3(H) = \sum_{1 \leq i < j < k \leq d_c} S_3(\{h_i, h_j, h_k\}),
		\label{eq:s3t}
\end{align}
\noindent and the term $S^c_3(H)$ indicates the summation of binary Hamming weight 3 associated to all possible couples of non-null coefficients, i.e.,

\begin{align}
	S^c_3(H) = \sum_{1 \leq a < b \leq d_c} S_3(\{h_a, h_b\}),
		\label{eq:s3c}
\end{align}
\\

\noindent  \textbf{Proof:} Let us consider a triplet $\{h_i, h_j, h_k\}$ of coefficients of a parity check of degree 3. The set $\mathcal{C}(x_i,x_j,x_k)$ of Hamming weight 3 triplets $(x_i,x_j,x_k)$ verifying $x_ih_i + x_jh_j + x_kh_k = 0$ can be partitioned in four disjoint sets: $\mathcal{C}^{j,k}_i = (0,x_j,x_k)_{x_j \neq 0, x_k \neq 0}$, $\mathcal{C}^{i,k}_j = (x_i,0,x_k)_{x_i  \neq 0, x_k \neq 0}$, $\mathcal{C}^{i,j}_k = (x_i, x_j ,0)_{x_i  \neq 0, x_j \neq 0}$ and $\mathcal{C}^{i,j,k} = (x_i,x_j,x_k)_{x_i  \neq 0, x_j \neq 0, x_k \neq 0}$. One can note that the number of elements of $\mathcal{C}^{i,j}_k$ is independent of $k$ and is equal to $|\mathcal{C}^{i,j}_k| = S_3(\{h_i, h_j\})$. Thus $S_3(\{h_i, h_j, h_k\}) = |\mathcal{C}^{i,j,k}| + S_3({h_i, h_j}) + S_3({h_i, h_k}) + S_3({h_i, h_k})$. According to (\ref{eq:s3t}),  $S_3^t$ is thus equal to

\begin{align*}
	S^t_3(H) = &\sum_{1 \leq i < j < k \leq d_c}  |\mathcal{C}^{i,j,k}| \\
	           &+ S_3(\{h_i, h_j\}) + S_3(\{h_i, h_k\}) + S_3(\{h_i, h_k\}). 
		\label{eq:s3tn}
\end{align*}

\noindent Since a given couple $\{h_a, h_b\}$ appears exactly $(d_c - 2)$ times in the right part of 
(\ref{eq:s3tn}). Thus, $S^t_3(H)$ is equal to the number of Hamming weight 3 codeword with exactly 3 non-null GF($q$) symbols plus $(d_c - 2)$ times the number of Hamming weight 3 codeword with exactly 2 non-null GF($q$) symbols, i.e, $S^c(H)$. 
\noindent Thus, $S^t_3(H) - (d_c - 2)S^c_3(H)$ gives the number of Hamming weight 3 codewords with 3 non null GF($q$) symbols, while $S^c_3(H)$ gives the number of Hamming weight 3 codewords with exactly two non-null GF($q$) symbols. Thus, adding those two terms gives $S_3(H)$, the total number of Hamming weight 3 codewords $\blacksquare$\\

\noindent \textbf{Property 1}  Let $x = \alpha^a$ an element of GF($2^m$), then $W(x) = 1$ is equivalent to $ 0 \leq a < m$. In others words, the binary representation of $x$ contains exactly one non null value (the binary Hamming weight of $x$ is equal to 1, or $S^x[X] = X^1$)" is equivalent to the property $0 \leq a < m$. For example, if GF($2^3$) is defined by $P_3[X] = 1 + X + X^3$, then $\alpha^0 = (1,0,0)$, $\alpha^1 = (0,1,0)$, $\alpha^2 = (0,0,1)$ while $\alpha^3 = (1,1,0)$.\\

\noindent \textbf{Theorem 1:} Let $H = \{\alpha^{a_i}\}_{i=1, \ldots, d_c}$ be a set of $d_c$ non null coefficients in GF($2^m$)$^{d_c}$, then, if $m > 2$, $S_2(H) = 0$ is equivalent to 

\begin{equation}
		\forall i, j  \in \{1, 2, \ldots d_c\}^{2}, i \neq j \Rightarrow |a_j - a_i|_{q-1} \geq m,
		\label{eq:s_2null}	
\end{equation}

\noindent where $|a|_{q-1}$ represents $\min(|a|, |q - 1 - a|)$.\\

\noindent \textbf{Proof:}  
Let us first prove the equivalence for a check node of degree $d_c = 2$ with the set of coefficients $\{h_1, h_2\}$, where $h_1 = \alpha^{a_1}$ and $h_2 = \alpha^{a_2}$ and $a_2 \geq a_1$. Since multiplying the coefficients of the check node by $\alpha^{-a_1} $ does not change the code,  $h_1$ can be set to $h_1 = \alpha^{0}$ and $h_2$ can be set to $\alpha^a$; with $a = a_2 - a_1$. The $q-1$ non null solutions of the parity check equation are thus $(x^b_1 =\alpha^{b+a}, x^b_2 = \alpha^b)$, $b = 0, 1, \ldots q-2$. In fact, $h_1x^b_1 + h_2x^b_2 = \alpha^0\alpha^{b+a} + \alpha^a\alpha^b = \alpha^{a+b} + \alpha^{a+b} = 0$. For a given $b$, the Hamming weight of the codeword $(x^b_1, x^b_2)$ is equal to $W(x^b_1) + W(x^b_2)$. According to property 1, we have $W(x^b_1) = 1$ equivalent to $0 \leq a + b \mod q-1 < m$ or equivalently 
\begin{equation}
(0 \leq  b  < m-a) \text{ or } (q-1-a \leq  b  < q-1).
\label{const1} 
\end{equation}
\noindent Similarly, $W(x^b_2)=1$ is equivalent to 

\begin{equation}
0 \leq b < m. 
\label{const2} 
\end{equation}

\noindent Thus, according to lemma 1, $W(x^b_1) + W(x^b_2) = 2 \Rightarrow W(x^b_1) = 1$ and $W(x^b_2) = 1$, or equivalently, there exists a value of $b$ that satisfies simultaneously (\ref{const1}) and (\ref{const2}). There is a solution if and only if $0 \leq m-1-a$ or $q-1-a \leq m - 1$. If $m>2$, the second inequality is never fulfilled and the existence of solution is given by $a \leq m-1$. Reciprocally, for $m>2$, if $a > m-1$, then $W(x^b_1) + W(x^b_2)$ is always strictly greater than 2. The general case can be proven by using lemma 2 $\blacksquare$\\

\noindent \textbf{Corollary}: $d_c \leq \frac{2^m}{m}$ is a necessary and sufficient condition for the existence of a set $H$ of $m$ non-null coefficients of $GF(q)$ so that $S_2(H)=0$.\\

\noindent \textbf{Proof:} $H = \{\alpha^0, \alpha^m, \alpha^{2m}, ..., \alpha^{(d_c-1)m}\}$ verifies (\ref{eq:s_2null}) if and only if $(d_c-1)m \leq q - m$ $\blacksquare$\\

The above properties are now used to find optimal (or optimized) sets of parity check equation coefficients.
	
\section{Determination of optimal coefficients}
\label{sec:complexity}

The objective of this paper is to find for several values of $d_c$ and Galois Field GF($q$) the sets of coefficients that minimize $S_3(H)$ with $S_2(H) = 0$. The design objective can be formalized as

\begin{equation}
		H^{opt} =  \arg \min_{H \in {\text{GF}(q)^*}^{d_c}} \{ S_3(H) / S_2(H) = 0 \},
		\label{eq:optimize}	
\end{equation}
 
\noindent where GF($q$)$^*$ is the set of non-null elements of GF($q$). Moreover, in the case where $S_3(H^{opt}) = 0$, the design objective is modified as

\begin{equation}
		H^{opt} =  \arg \min_{H \in {\text{GF}(q)^*}^{d_c}} \{ S_4(H) / (S_3(H) = 0, S_2(H) = 0) \}.
		\label{eq:optimize}	
\end{equation}
 
This section is divided in 3 sub-sections. First, we explicitly describe how to compute efficiently $S_3(H)$ (or $S_4(H)$) given the set $H$. Then, we determine $\xi_m(d_c)$, the number of sets $H$ verifying $S_2(H) = 0$ as a function of the degree $d_c$ of the parity check and the order $q=2^m$ of the Galois Field. Finally, when $\xi_m(d_c)$ is too high for an exhaustive search, we propose a heuristic to find good sets of coefficients.

\subsection{Determination of $S_3(H)$}

In the sequel, we propose an efficient method to compute the value of $S_3(H)$ for several sets of coefficients $H$. The first step is to compute tables $T_2$ and $T_3$. Table $T_2$ is a table of size $(q-1)$ defined as $T_2(a)=S_3(\{\alpha^0, \alpha^{a}\})$ for $a = 0, \ldots q-2)$. Table $T_3$ is a table of size $(q-1)\times(q-1)$ defined as $T_3(a, b) = S_3(\{\alpha^0, \alpha^{a}, \alpha^{b}\})$, $(a,b) \in \{0, 1, \ldots, q-2\}^2$. For a given couple $(a,b)$, the computation of $T_3(a,b)$ requires $q^2$ operations using algorithm 1. The determination of the whole table has thus a global complexity of $q^4$. This complexity is high ($10^{12}$ for $q = 1024$) but since it is processed only once for all sets of coefficients $H$ for a given Galois Field order $q$, it is still feasible.

Once tables $T_2$ and $T_3$ are generated, $S_3(H)$ is obtained thanks to the computation of $S^c_3$ and $S^t_3$, as described in algorithm \ref{algo:S_3}.

\begin{algorithm}
 {
 \label{algo:S_3}
 \KwData{Initial set of coefficients $H = \{\alpha^{a_i}\}_{i=0,\ldots, d_c - 1}$}
 \KwResult{$S_3(H)$}
 $S_3^c = 0$; $S_3^t = 0$\\
 \For{$i=0, \ldots d_c-2$}
	 {
	 \For{$j=i+1, \ldots d_c-1$}
				 { 
						$S_3^c = S_3^c  + T_2(a_j - a_i)$;\\
				 }
	 }
 \For{$i=0, \ldots d_c-3$}
	 {
	 \For{$j=i+1, \ldots d_c-2$}
				 { 
				 \For{$k=j+1, \ldots d_c-1$}
						{
						$S_3^t = S_3^t  + T_3( a_j - a_i, a_k - a_i)$;
            }
				 }
		}
		$S_3(H) = S_3^t - (d_c - 3)S_3^c$\\	  
 }
\caption{Determination of $S_3(H)$}
\end{algorithm}

To summarize, once tables $T_2$ and $T_3$ computed, the additional computational cost to determine $S_3(H)$ for a parity check node of degree $d_c$ is independent of the Galois Field order $q$ and requires exactly $C(d_c) = \binom{2}{d_c} + \binom{3}{d_c} =  \frac{1}{6}({d_c}^3 - d_c)$ table accesses and add operations. This low complexity (for $d_c= 20$, $C(20) = 1330$) permits to test rapidly a large number of potential sets of coefficients. 

\subsection{Determination of $H_3^{opt}$}

Let us consider a check node of degree $d_c \leq \frac{2^m}{m}$. The enumeration of all sets of coefficients verifying $S_2(H) = 0$ required to determine $H^{opt}$ can be performed thanks to algorithm \ref{algo:enum}.

\begin{algorithm}
 \label{algo:enum}
 \KwData{Parity check degree $d_c$, $ m = \log_2(q)$}
 \KwResult{$H ^{opt}$}
  $s_3^{opt} = +\infty$ ; $s_4^{opt} = +\infty$\\
  \For{$a_1 = m, \ldots, q -1 - m(d_c-1)$}
	 {
	 \For{$a_2 = a_1 + m, \ldots, q -1 - m(d_c-2)$}
				{
				 \ldots \\
				\For{$a_i = a_{i-1} + m, \ldots, q - 1 -m(d_c-i)$}
				    {
							\ldots \\
							\For{$a_{d_c-1} = a_{d_c-2}+ m, \ldots, q - 1 - m $}
								{
									$ H = \{\alpha^{0}, \alpha^{a_1}, \ldots, \alpha^{a_i}, \ldots, \alpha^{a_{d_c - 1}}\}$ \\
									Compute $s_3 = S_3(H)$ thanks to algorithm 2 \\
									\If {$s_3 < s_3^{opt}$}
										{$H^{opt} = H$ \\
										$s_3^{opt} = s_3$\\
										}
									\If {$s_3 = 0$}
											{
 											Compute $s_4 = S_4(H)$ thanks to algorithm 1 \\
												\If {$s_4 < s_4^{opt}$}
														{ 
														$H^{opt} = H$;\\
														$s_4^{opt} = s_4$; \\
											      }
											}
								}
					  }
				}
	}

\caption{Determination of $H^{opt}$}
\end{algorithm}

It is useful to compute the exact number of configurations to be tested in order to explore all possible sets of coefficients leading to $S_2(H) = 0$. Let us first introduce the set $\Gamma_m(p, n)$ defined as the set of $p$-tuplet of integers $({a(1)}, {a(2)}, \ldots, {a(p)}) \in \{0, 1, \ldots n-1 \}^p$ verifying the following constraint

\begin{equation}
    a(i+1) - a(i) \geq m, \: i=1, 2, \ldots, p-1
\label{eq:pair}
\end{equation}

Appendix II gives a method inspired from the Pascal's Triangle \cite{pascal} to compute the cardinality  $\gamma_m(p, n)$  of  the set $\Gamma_m(p, n)$.

Let $\xi_m(d_c)$ be the number of sets of coefficients in a check node of degree $d_c$ over GF($2^m$) that verifies the condition of theorem 1 (i.e. that gives a minimum Hamming weight 3 for its equivalent binary code). The first coefficient can be always $h_1 = \alpha^0$, since the multiplication of all coefficients by the same constant value does not change the code \cite{Poulliat2006}. Once $\alpha^0$ is selected, $\{\alpha^1, \alpha^2, \ldots, \alpha^{m-1}\}$ and $\{\alpha^{q-m}, \ldots, \alpha^{q-3}, \alpha^{q-2}\}$ are removed in order to respect theorem 1, i.e.,  every pair of coefficients of the check node should have their logarithms separated by at least $m$ modulo $q-1$. Thus, there are still $p = d_c-1$ points to be placed among $2^m -1 - (2m-1)= 2^m -2m$ values (see Fig. \ref{fig:random_sampling}.a), and thus

	\begin{equation}
		\xi_m(d_c) = \gamma_m(d_c-1, 2^m - 2m).
		\label{number_of_configuration}
	\end{equation}

\begin{figure}[h!]
	\centering
	\includegraphics[width=9cm]{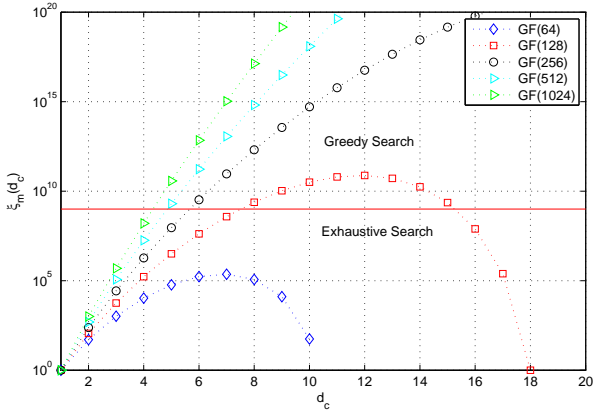}
	\caption{Number $\xi(d_c, m)$ of sets of coefficients for $d_c \leq 20$ over GF(64) up to GF(1024)}
	\label{fig:nb_configuration}
\end{figure}

For example, according to Table \ref{tab:phi_GF16}, there is exactly $\xi_5(4) = \gamma_5(3, 22) = 364$, i.e., there is 364 sets of coefficients, with the first one equal to $\alpha^0$, that lead to a Hamming distance of 3 for a check node of degree 4 over GF($2^m=32$) (coefficients are supposed to be sorted in increasing order of their logarithm). It is thus easy to generate these 364 solutions in order to keep the ones leading to the minimum multiplicity of weight 3 codewords (i.e., the minimum of $S_3(H)$). One should note that $\xi_m(d_c)$ is exactly the number of time that a set $H$ is tested in algorithm \ref{algo:enum}. Fig. \ref{fig:nb_configuration} shows the number of configurations $\xi_m(d_c)$ for $m$ equal to 6 (GF(64)) to 10 (GF(1024)) and $d_c$ varying from 1 to 20. 

Note that $\xi_8(20) = 2.39 \times 10^{22}$ (not shown in Fig. \ref{fig:nb_configuration}), which is a far too high number for an exhaustive search. In this paper, the limit for an exhaustive exploration to determine $H^{opt}$ is set to configurations where $\xi_m(d_c) < 10^9$. When $\xi_m(d_c) \geq 10^9$, a heuristic search should be used to find good sets of coefficients.

\subsection{Heuristic search of coefficients} \label{sec:result}

When the value of $\xi_m(d_c)$ is too high for an exhaustive exploration, a heuristic search should be used. In this paper, we propose a basic but effective method. It is based on a greedy search repeated several times, each attempt starting from an initial state taken randomly. Let $N_g$ be the number of attempt, $H^{0,k}$ the $k^{th}$ random initial set of coefficients, $\tilde{H}^{0,k} = G(H^{0,k})$ the final state obtained when a greedy algorithm is applied on $H^{0,k}$. The optimized solution $H_3^{f}$ is taken as

\begin{equation}
\label{eq:greedy}
H_3^f= \arg \min \{S_3( \tilde{H}^{0,k}), k = 1 \ldots N_g\}.
\end{equation}

Let us describe in more details the method to draw the $H^{0,k}$ and the greedy algorithm. 
							
\subsubsection{Method to generate initial sets of coefficients}

The generation of the initial set should be unbiased, i.e., any set of coefficients should have the same probability $P = \frac{1}{\xi_m(d_c)}$ of being chosen. This requirement can be achieved by a step by step generation process. In the sequel, the index $k$ is omitted for clarity. 

\begin{figure}
	\centering
	\includegraphics[width=9cm]{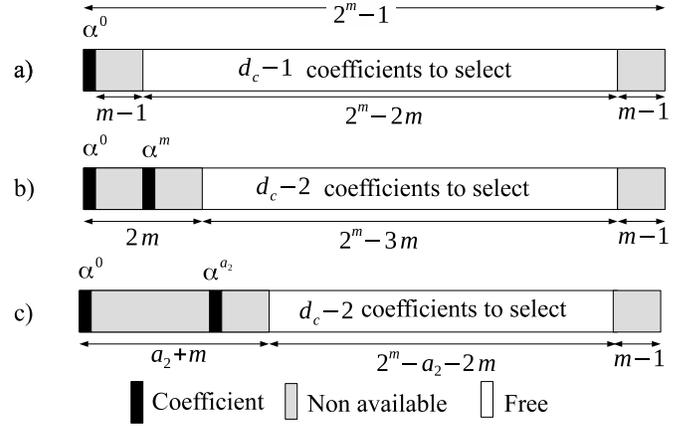}
	\caption{Illustration of the random coefficients selection process.}
	\label{fig:random_sampling}
\end{figure}

The first element $h^0_1$ of $H^0$ is always $h^0_1 = \alpha^0$. Then, the smallest (in the sense of logarithm over GF(q)) next element is $h^0_2 = \alpha^m$ ($a_2 = m$) according to theorem 1. In that case, there are still $d_c -2$ coefficients to be drawn among $2^m - 3m$ positions, as shown if Fig. \ref{fig:random_sampling}.b. The number of elements is thus $\gamma_m(d_c-2, 2^m - 3m)$ possibilities. If the next chosen element is $a_2 > m$, as shown in \ref{fig:random_sampling}.c, there is still $d_c - 2$ coefficients to be drawn among $2^m - a_2 - 2m$, and thus $\gamma_m(d_c -2, 2^m - a_2 - 2m)$ possibilities. In order to draw a set of coefficients randomly, we should have, for the second coefficient:

\begin{equation}
\label{eq:random2}
\text{Prob}( h^0_2 = \alpha^{a_2}) = \frac{\gamma_m(d_c-2, 2^m - a_2 - 2m)}{\gamma_m(d_c-1, 2^m - 2m)}.
\end{equation}

\noindent One should note that the sum of the probability $\text{Prob}(h^0_2 = \alpha^{a_2})$ for all values of ${a_2}$ is equal to 1 according to (\ref{eq:sum}). For the third element (and the fourth up to the last one), the same method can be applied, leading to the general formula to generate the $j^{th}$ coefficients $a_j$ knowing that the previous coefficient is $a_{j-1}$, $a_j>a_{j-1}$ is given by

\begin{align}
\label{eq:random_a}
\text{Prob}( h^0_j = \alpha^{a_j}/h^0_{j-1} = \alpha^{a_{j-1}}) = \nonumber \\
\frac{\gamma_m(d_c-j, 2^m - 2m - {a_j})}{\gamma_m(d_c-j + 1, 2^m - 2m - a_{j-1})}.
\end{align}

To conclude, the generation of uniformly distributed sets of coefficients reduces to a Markov process where probability of transition at a given stage is given by (\ref{eq:random_a}). Finally, the generation of an initial set of coefficients using random values can be enriched by a method proposed by one of the reviewers' paper. They suggested to insert a new coefficient in an already optimized set of coefficients of lower degree to create an a priori good seed value for the greedy optimization process. For high GF order ($q = 512$, 1024) and $d_c$ values, this method sometimes allows to reduce the value of $S_3$ by a few units.

\subsubsection{Proposed greedy algorithm}

The initial set of $H$ coefficient is $H = H^0$, then all possible values for the second coefficient $h_2$ verifying $h_2 = \alpha^{a_2}$, with $m \leq a_2 \leq a_3-m$ are tested. This limited search space guaranties that $S_2(H) = 0$ (see Theorem 1). The value of $a_2$ that minimizes $S_3$ is selected to generate the new set of coefficients $H$. Then the same process is applied on the third coefficient (with $a_2 + m \leq a_3 \leq a_4-m$) up to the $d_c^{th}$ coefficient. The whole process is started again until no more improvement is obtained. The algorithm is given in details in algorithm \ref{algo2}. Note that when $l = d_c$, $l+1$ goes back to 1, and thus, $a_{d_c+1}-m =  - m \mod 2^m - 1 = 2^m - m$. One should note that many more sophisticated and efficient algorithms can be imagined. Nevertheless, repeated many times from random initial states, the overall search method is effective.

\begin{algorithm}
 \KwData{Initial set of coefficients $H^0 = \{h_i = \alpha^{a_i}\}_{i=1,\ldots d_c}$}
 \KwResult{Final set of coefficients $\tilde{H} = G(H^0)$}
 $ \tilde{H} = H^0$ \\
 $s_3^{opt}  = S_3(\tilde{H})$; \\
 Improved = true; \\
 \While{Improved}
{
	Improved = false;
 \For{$i=1, \ldots d_c-1$}
	 {
    $H = \tilde{H}$;   \\
    \For{$b = a_{i-1} + m$, ..., $a_{i+1}-m$}
		  		 {
							 $a_i = b$; \: \: ($H = \{\alpha^{a_0}, \ldots, \alpha^{a_{i-1}}, \alpha^{a_{b}}, \alpha^{a_{i+1}}, \ldots, \alpha^{a_{d_c-1}}\}$) \\
 							  Compute $s_3 = S_3(H)$ thanks to algorithm 2; 			\\
								\If {$s_3 < s_3^{opt}$}
											{
											$\tilde{H} = H$;\\
											$s_3^{opt} = s_3$; \\
											Improved = true; \\
										  }
										 
				  }
     }
}
\caption{Greedy algorithm to compute $\tilde{H} = G(H^0)$.}
\label{algo2}
\end{algorithm}
 										
\begin{figure}
	\centering
	\includegraphics[width=9cm]{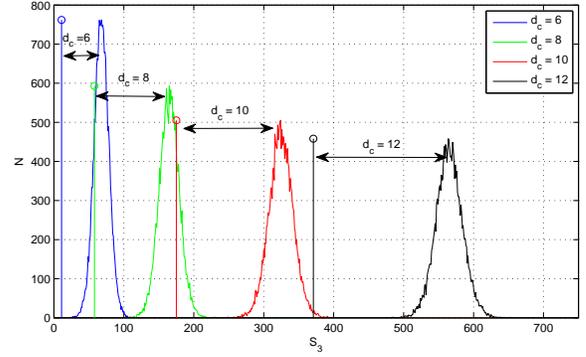}
	\caption{Histogram of $S_3(H^0)$ and the best value $S^{opt}_3$ over GF(256) for several values of check node degree $d_c$}
	\label{fig:histogramme}
\end{figure}

\begin{figure}
	\centering
	\includegraphics[width=9cm]{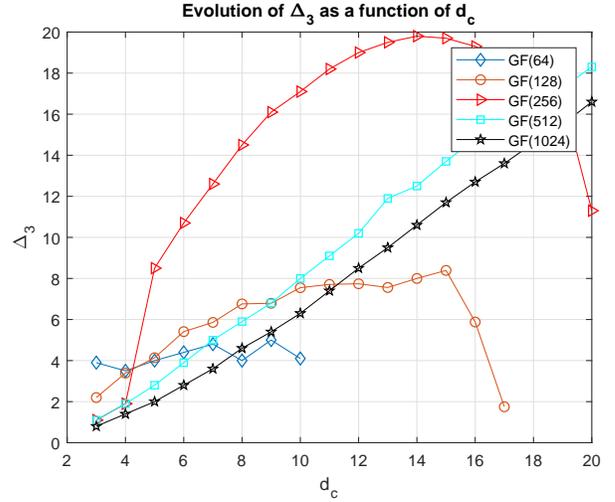}
	\caption{Value of $\Delta_3$ as a function of $d_c$ for GF(64) up to GF(1024)}
	\label{fig:Delta_3}
\end{figure}

\begin{figure}
	\centering
	\includegraphics[width=9cm]{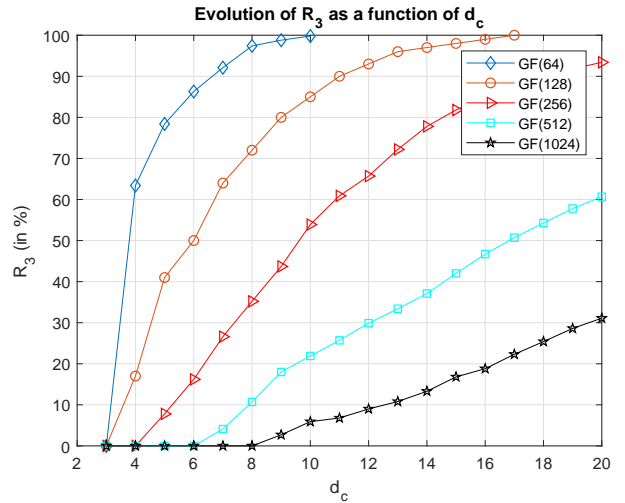}
	\caption{Value of $R_3$ as a function of $d_c$ for GF(128) up to GF(1024)}
	\label{fig:M_3}
\end{figure}

Fig. \ref{fig:histogramme} shows the histogram of $S_3(H^0)$ obtained with $N=20,000$ draws as well as the best value found for $d_c = 6, 8, 10$ and 12  over GF(256). In order to evaluate how far is the best found solution $S_3^f$ compared to the average value of $S_3(H^0)$, we use the two following metrics

\begin{align}
\Delta_3 &= \frac{M_3 - S_3^f}{\sigma_3}  \\
 R_3 &= \frac{S_3^f}{M_3} \times 100 \:\: (\textrm{in} \: \%) 
\label{eq:dist_sigma}
\end{align}

\noindent where $M_3$ and $\sigma_3$ are respectively the mean and the standard deviation of $S_3(H^0)$ for $H^0$ satisfying $S_2(H^0) = 0$. The first metric $\Delta_3$ measures how far is the found value relatively to the "gaussian like shape" distribution of $S_3(H^0)$ while the second metric indicates the relative gain, in \%, compared to the mean value $M_3$. Fig. \ref{fig:Delta_3} and Fig. \ref{fig:M_3} show the evolution of $\Delta_3$ and $M_3$ for several values of $d_c$ and GF($q$) order. 
Values of $H_3^f$, $M_3$, $\sigma_3$ and the corresponding set of coefficients are given for GF(64) up to GF(1024) in Appendix II.

\section{Conclusion}


In this paper, we have generalized the method proposed by Poulliat et
al. for the determining the optimal Galois Field coefficients of a
Non-Binary LDPC parity check code based on the binary image of the
code. An algorithm with a complexity in $\mathbb{O}(d_c^3)$ has been
proposed to determine the number $S_3(H)$ of codewords of binary Hamming
weight 3 of a parity check of degree $d_c$ over GF($q$). The low
computational complexity of the algorithm opens exploration to new regions
of the design space, i.e. check node degree $d_c$ greater than 4 and
high order Galois Field (up to GF(1024)) by an exhaustive search.  A new
greedy search algorithm has also been proposed to find good solutions
when the number of sets of coefficients is too high for an exhaustive
search. Tables of sets of coefficients are given for values of $d_c$
between 4 and 20 and GF($q$) order varying from $q=64$ to $q=1024$. For each set of
coefficients, the best found value $S_3^f(H)$ is compared with the
distribution of $S_3(H)$ obtained by taking randomly the coefficients of
$H$. In some cases, $S_3^f(H)$ can be at a distance to the mean value of
$S_3(H)$ greater than 10 times the standard deviation of the
distribution. The proposed sets of coefficients can effectively replace
the random selection of coefficients often used in NB-LDPC construction
over high order Galois Field, and thus helps the construction of new
generations of NB-LDPC codes with better decoding performance.

\section*{Acknowledgment}
The author would like to thank C\'edric Marchand who pointed out some errors in the preliminary draft of the paper. He also thanks the reviewers who helped to improve the paper and suggested good ideas.

\section*{Appendix I}

It is useful to compute the exact number of configurations to be tested in order to explore all possible sets of coefficients leading to $S_2(H) = 0$. To do so, we use a method inspired from the Pascal' Triangle method \cite{pascal}. 

Let $\Gamma_m(p, n)$ be the set of $p$-tuplet of integer $({a(1)}, {a(2)}, \ldots, {a(p)})$ verifying the following two constraints

\begin{equation}
    a(i) \in \{0, 1, \ldots n-1 \}, \: i = 1, 2, \ldots, p
\label{eq:card}
\end{equation}

\noindent and

\begin{equation}
    a(i+1) - a(i) \geq m, \: i=1, 2, \ldots, p-1
\label{eq:pair}
\end{equation}

\noindent The cardinality $|\Gamma|$ of set $\Gamma$ will be denoted as $\gamma = |\Gamma|$. According to this definition, $\Gamma_6(2, 8)$ is equal to $\Gamma_6(2, 8) = \{(0, 6), (0, 7), (1, 7) \}$ and the cardinality of $\Gamma_6(2, 8)$ is $\gamma_6(2, 8) = 3$. \\

\noindent \textbf{Case $p = 1$}: When $p = 1$, then only constraint (\ref{eq:card}) can be applied and thus $\Gamma_m(1, n) = \{0, 1, \ldots, {n}\}$ and $\gamma_m(1, n) = n$.\\

\noindent \textbf{Case} $p = 2$: When $p = 2$, if $n \leq m$, there is no solution, thus $\Gamma_m(2, n) = \emptyset$ and $\gamma_m(2, n) = 0$. 
If $n = m+1$, there is a unique solution $\Gamma_m(2, m+1) = \{(0, m)\}$ and thus $\gamma_m(2, m+1)=1$.\\
If $n = m+2$, there are 3 possible solutions: $\Gamma_m(2, m+1) = \{(0, m), (0, {m+1}), (1, {m+1})\}$ and $\gamma_m(2, m+1) = 3$.\\
If $n = m+3$, there are 6 elements $\Gamma_m(2, m+3)$. In fact, the elements of $\Gamma_m(2, m+2)$ belongs also to $\Gamma_m(2, m+3)$. The additional elements are the 3 couples $(0, {m+2})$, $(1, {m+2})$ and $(2, {m+2})$. These 3 couples can be represented by $\{\Gamma_m(1, n-m)\parallel{m+2}\}$, where $\{\Gamma||x\}$ means the set obtained by concatenating $x$ on the right to all elements of $\Gamma$. In other words, $\Gamma_m(2, m+3) = \Gamma_m(2, m+2) \cup \{\Gamma_m(1, n-m)||{m+2} \}$, and thus, $\gamma_m(2, m+3) = \gamma_m(2, m+2) + \gamma_m(1, n-m)$.\\

\noindent \textbf{General Case}: In the general case, $\Gamma_m(2, n) = \Gamma_m(2, n-1) \cup \{\Gamma_m(1, n-m)||{n-1}\}$ and thus

	\begin{equation}
		\gamma_m(2, n) = \gamma_m(2, n-1) + \gamma_m(1, n-m).
		\label{phi2_recursion}
	\end{equation}
	
\noindent In (\ref{phi2_recursion}), we recognize the structure of the Pascal's triangle binomial construction, and thus
	\begin{equation}
		\gamma_m(2, n) = \binom{2}{n-m+1} = \frac{(n-m+1)(n-m)}{2}.
		\label{phi2_formula}
	\end{equation}

In the general case, $\Gamma_m(p, n)$ and $\gamma_m(p, n)$ can be determined by a double recursive equation. First, let us assume that $\Gamma_m(p', n')$ are known for all couples $(p' < p, n' \in \mathbb{N})$ and $(p, n' < n)$. Then, $\Gamma_m(p, n)$ can be generated as
 
	\begin{equation}
		\Gamma_m(p, n) = \Gamma_m(p, n-1) \cup \{\Gamma_m(p-1, n-m)||{n-1} \}.	
		\label{phip_recursion}
	\end{equation}

\noindent This equality gives
	\begin{equation}
		\gamma_m(p, n) = \gamma_m(p, n-1) + \gamma_m(p-1, n-m).
		\label{phip_recursion_card}
	\end{equation}
		
The derivation of the exact value of $\gamma_m(p, n)$ is out of the scope of this paper. We can nevertheless derive from the recursion method that

\begin{equation}
\gamma_m(p, n) = \sum_{k=1}^{n}{\gamma_m(p-1, k-m)}.
\label{eq:sum}
\end{equation}

The important point is that the exact number of configurations can be, in practice, determined. As an example, Tab. \ref{tab:phi_GF16} gives the first values of $\gamma_5(p, n)$ for $p \leq 5$ and $n \leq 22$

\begin{table*}
	\centering
		\begin{tabular}{|r|r r r r r r r r r r r r r r r r r r|} \hline
 value of $n$ & $n\leq 5$ & 6 & 7 & 8 & 9 & 10 & 11 & 12 & 13 & 14 & 15 & 16 & 17 & 18 & 19 & 20 & 21 & 22 \\ \hline
	 $p=1$ & $n$ & 6 & 7 & 8 & 9 & 10 & 11 & 12 & 13 & 14 & 15 & 16 & 17 & 18 & 19 & 20 & 21 & 22 \\ 
	 $p=2$ & 0 & 1 & 3 & 6 & 10& 15 & 21 & 28 & 36 & 45 & 55 & 66 & 78 & 91 & 105 & 120 & 136 & 153 \\
	 $p=3$ & 0 & 0 & 0 & 0 & 0 & 0 & 1 & 4 & 10 & 20 & 35 & 56 & 84 & 120& 165 & 220 & 286 & 364 \\
	 $p=4$ & 0 & 0 & 0 & 0 & 0 & 0 & 0 & 0 & 0 & 0 & 0 & 1 & 5 & 15& 35 & 70 & 126 & 210 \\
	 $p=5$ & 0 & 0 & 0 & 0 & 0 & 0 & 0 & 0 & 0 & 0 & 0 & 0 & 0 & 0& 0 & 0 & 1 & 6 \\
			\hline
 		\end{tabular}
	\caption{Values of $\gamma_5(p, n)$}
	\label{tab:phi_GF16}
\end{table*}

\section*{Appendix II} \label{sec:result}

In this appendix, we give the results obtained by the proposed methods in
order to help the construct optimal, or almost optimal, NB parity
check codes. Remind that multiplying a set of coefficients by a constant factor does not change the
code. For example $H = \{\alpha^0, \alpha^{9},    \alpha^{22},
\alpha^{37}\}$ over GF(64) gives the same code as $H' = H\alpha^{54} =
\{\alpha^{54}, \alpha^{63},    \alpha^{76},    \alpha^{91}\} =
\{\alpha^{54}, \alpha^{0},    \alpha^{13},    \alpha^{28}\}$. After
reordering of the coefficients, $H'$ is equal to $H'= \{\alpha^{0},
\alpha^{13},    \alpha^{28},    \alpha^{54}\}$. Since the parity check
generated by $H$, $H'=H\alpha^{54}$, $H''=H\alpha^{41}$ and
$H'''=H\alpha^{26}$ are all equal, only the set of coefficients that
minimizes the value of $a_2$ will be given to represent the equivalent
set of coefficients through a multiplicative factor. When distinct
optimal solutions exist for a given configuration of $d_c$ and GF($q$),
those solutions are enumerated.


\begin{table*}[h!]
\begin{minipage}{16cm}
		\begin{tabular}{|r|r r| r r r r | l |} \hline
%
$d_c$ &	 $S_3^f$	& $S_4^f$ & $M_3$ & $\sigma_3$ & $\Delta_3$ & $R_3$ (\%) & GF(64)\\			\hline 
 
 3	 &      0	 &  68   &  12.0    &  3.1    &    3.9     &  0 \%  &$\{1,	16,	42\}$  \\
 4\footnote{This set of coefficients was initially proposed in \cite{website2003} and \cite{Poulliat2006}}
	 & $^*$20	 &  206  &  31.5 &  3.3 &   3.5 &  63.4  \% & $\{0,	9,	22,	37\}$ \\
 5\footnote{In \cite{website2003}, a list of 77 of sets of
coefficients are given for $d_c = 5$ over GF(64). In this list, some
sets of coefficients have $S_2(H) > 0$. The best proposed one is $H =
\{\alpha^1, \alpha^7, \alpha^{36}, \alpha^{58}\}$ with $S_2(H) = 0$ and
$S_3(H) = 57$.}
  	 & $^*$51	 & 500 & 65.0  &  3.5 &   4.0 &  78.4  \% & $\{0, 7, 18, 44, 53\}$  \\
 6	 & $^*$100 &  1020 &115.9  &  3.6 &   4.4 &  86.3  \% & $\{0, 6, 13, 20, 46, 55\}$ \\
 7	 & $^*$173 &  1890 &187.9  &  3.1 &   4.8 &  92.1  \% & $\{0,  6, 13, 21, 28, 44, 54\}$\\
 8	 & $^*$276 &  3211 &283.3  &  1.7 &   4.0 &  97.4  \% & $\{0,  6, 13, 21, 28, 36, 44, 54\}$\\
 9	 & $^*$402 &  5196 &406.8  &  1.0 &   5.0 &  98.8  \% & $\{0, 6, 14, 21, 27, 35, 42, 48, 56\}$ \\
 10	 & $^*$560 &  7995 &560.9  &  0.2 &   4.1 &  99.8	 \% & $\{0,  6, 12, 18, 24, 30, 37, 44, 50, 56\}$ \\
			\hline
\end{tabular}
\end{minipage}
\caption{List of optimal coefficient's exponents $\{a_i\}_{i=1, \ldots d_c}$ for GF(64). The symbol $^*$ indicates that the value of $S_3^f$ is equal to $S_3^{opt}$.}
	\label{tab:gf_inf_64}
\end{table*}

\begin{table*}[h!]
	\centering
   \begin{minipage}{16cm}
		\begin{tabular}{|r|r r| r r r r | l |} \hline
$d_c$ &	 $S_3^f$	&  $S_4^f$	& $M_3$ & $\sigma_3$ & $\Delta_3$ & $R_3$ (\%) & GF(128), \\ \hline
 3   &    $^*$0  &  52 &   9.0   &  4.0       &   2.2  &  0 \%        &  $\{0, 15, 53\}$ ; $\{0, 15, 54\}$; $\{0, 73, 88\}$ ; $\{0, 74, 89\}$ ; \\
    &        &     &         &            &        &              &   $\{0, 38, 112\}$ ; $\{0, 39, 112\}$ \\
 4\footnote{In \cite{Poulliat2006}, the best given sets of coefficients have $S_3(H) = 5$ }
 	  & $^*$4	    & 244 &23.3 &  5.7    & 3.39 & 17 \% & $\{0, 12, 84, 101\}$\\
 5  & $^*$20  &   552 &49.0 &  7.0    & 4.14& 41 \% & $\{0, 11, 55, 84, 101\}$ \\
 6  & $^*$44  &   1111 &87.8 &  8.1    & 5.41& 50 \% &  $\{0, 9, 21, 60, 94, 112\}$ \\
 7  & $^*$92  &   1985 &143.0 & 8.7    & 5.86& 64 \% &  $\{0, 7, 24, 39, 48, 60, 99\}$\\
 8  & $^*$157 & 3283   &217.2 & 8.9    & 6.76& 72 \% &  $\{0, 7, 30, 37, 54, 69, 78, 90\}$\\
 9  & $^*$252 & 5159   &313.1 & 9.0    & 6.79& 80 \% &  $\{0, 7, 19, 30, 37, 54, 69, 78, 90\}$\\
 10 & 370     & 7739   & 433.4 & 8.4    & 7.55& 85 \% &  $\{0, 7, 38, 45, 59, 68, 75, 92, 107, 116\}$\\
 11 & 522     & 11206   &581.4 & 7.7    & 7.71& 90 \% &  $\{0, 7, 22, 30, 37, 48, 55, 69, 78, 89, 96\}$\\
 12 & 709     & 15759   &759.3 & 6.5    & 7.74& 93 \% &  $\{0, 7, 18, 25, 39, 48, 59, 66, 88, 97, 104, 119\}$\\
 13 & 928     & 21613   &969.6 & 5.5    & 7.56& 96 \% &  $\{0, 7, 17, 24, 38, 48, 58, 65, 72, 87, 96, 103, 118\}$\\
 14 & 1182    & 29067   &1215.6  & 4.2    & 8.00& 97 \% &  $\{0, 7, 14, 29, 38, 45, 55, 62, 69, 76, 86, 93, 107, 116\}$\\
 15 & 1473    & 38349   &1499.0  & 3.1    & 8.39& 98 \% &  $\{0, 7, 18, 25, 32, 39, 47, 54, 61, 69, 78, 89, 96, 103, 118\}$\\
 16 & $^*$1813& 49714   & 1823.0  & 1.7    & 5.88& 99 \% &  $\{0, 7, 17, 24, 31, 38, 47, 54, 61, 68, 75, 82, 89, 96, 103, 118\}$\\
 17 & $^*$2190& 63526   & 2190.7  & 0.4    & 1.75& 100 \% & $\{0, 7, 14, 21, 28, 35, 42, 49, 56, 63, 70, 77, 84, 91, 98, 105, $\\
    &        &     &         &            &        &              &   $119\}$\\
 18 & $^*$2604& 80073   & 2604.0  & 0.0    & 0.0 & 100 \% & $\{0, 7, 14, 21, 28, 35, 42, 49, 56, 63, 70, 77, 84, 91, 98, 105,$ \\
     &        &     &         &            &        &              &   $112, 119\}$\\
			\hline 
\end{tabular}
\end{minipage}
\caption{List of optimal or optimized ($N_g$ = 20,000) sets of coefficient's exponents $\{a_i\}_{i=1, \ldots d_c}$ for GF(128). The symbol $^*$ indicates that the value of $S_3^f$ is equal to $S_3^{opt}$.}
	\label{tab:gf_inf_128}
\end{table*}

\begin{table*}[h!]
	\centering
	  \begin{minipage}{16cm}
		\begin{tabular}{|r|r r| r r r r | l |} \hline
$d_c$ &	 $S_3^f$	& $S_4^f$	&$M_3$ & $\sigma_3$ & $\Delta_3$ & $R_3$ (\%) & GF(256) \\ \hline
 3	 & $^*$0	 & 36 &  7.3  &  4.4   &  1.7  &  0.0  \% 	& $\{0, 72, 80\}$ ; $\{0, 8, 183\}$ ; $\{0, 175, 247\}$  \\ 	
 4\footnote{This set is also given in \cite{Poulliat2006}}
     & $^*$0 &  156 &19.2   & 6.3  & 3.0   &   0.0 \% & $\{0, 8, 172, 183\}$ \\
 5	 & $^*$3 & 486 &38.6 & 4.2 & 8.5 & 7.8 \% & $\{0, 8, 66, 172, 180\}$  \\
 6	 & 11	 &  1014 &68.1 & 5.3 &10.7 &16.2 \% 	& $\{0, 8, 75, 83, 91, 150\}$  \\ 	
 7	 & 29	 &  1918 &109.2 & 6.3 &12.6 &26.6 \% 	& $\{0, 8, 76, 84, 92, 131, 150\}$\\
 8	 & 58	 &  3197 &164.5 & 7.3 &14.5 &35.2 \% 	& $\{0, 8, 36, 75, 83, 91, 129, 149\}$   \\
 9	 & 103 & 4952 &235.6 & 8.2 &16.1 &43.7 \%  & $\{0, 8, 37, 76, 84, 92, 130, 150, 234\}$  \\
 10	& 175	 &  7345 &324.7 & 8.8 &17.1 &53.9 \% 	&	$\{0, 8, 16, 54, 74, 139, 158, 179, 188, 215\}$  \\
 11	& 264	 &  10493 &433.6 & 9.3 &18.2 &60.9 \% 	&	$\{0, 8, 27, 92, 109, 131, 139, 169, 208, 216, 224\}$   \\
 12	& 371	 &  14689 &564.9 &10.2 &19.0 &65.7 \% 	&	$\{0, 8, 27, 39, 92, 109, 132, 140, 169, 208, 216, 224\}$	 \\
 13	& 522	 &  19781 &720.0 &10.2 &19.5 &72.2 \% 	&	$\{0, 8, 18, 38, 46, 65, 77, 130, 147, 170, 178, 207, 246\}$   	 	 \\
 14	& 701	 &  19781 &901.3 &10.1 &19.8 &77.8 \% 	&	$\{0, 8, 16, 42, 82, 90, 98, 107, 128, 136, 155, 167, 220, 237\}$     \\
 15	& 908  &  34212 &1110.3 &10.3 &19.7 &81.8 \% &	$\{0, 8, 29, 37, 76, 84, 92, 103, 123, 131, 150, 162, 192, 215, 233\}$ \\
 16	&1150	 &  43754 &1349.8 &10.6 &19.3 &84.9 \% &	$\{0, 8, 16, 34, 42, 79, 87, 95, 106, 126, 134, 153, 165, 196, 218, 236\}$	 	 \\
 17	&1426	 &  55347 &1621.3 &10.5 &18.7 &88.0 \% &	$\{0, 8, 45, 53, 61, 69, 77, 94, 102, 121, 133, 164, 186, 203, 221, 229,$\\
    &      &  &     &     &     &        & $237\}$\\
 18	&1737	 &  69408 &1926.9 &10.7 &17.8 &90.1 \% &	$\{0, 8, 19, 27, 35, 52, 60, 92, 100, 108, 116, 126, 147, 155, 173, 185,$ \\
    &      &  &     &     &     &        & $216, 237\}$ \\
 19	&2083 &  85992 &2268.5 &11.0 &16.8 &91.8 \% & $\{0, 8, 26, 39, 70, 91, 109, 117, 126, 134, 142, 161, 169, 183, 202, 210, $\\
    &     &  &     &     &     &        & $218, 226, 236\}$ \\
 20	&2473 &  105412&2648.4 &15.1 &11.3 &93.4 \% & $\{0, 8, 22, 30, 38, 52, 61, 75, 93, 101, 109, 117, 127, 147, 155, 174$\\
    &     &  &      &     &     &        & $186, 206, 216, 238\}$ \\
			\hline
\end{tabular}
\end{minipage}

\caption{List of optimal or optimized ($N_g$ = 20,000) sets of coefficient's exponents $\{a_i\}_{i=1, \ldots d_c}$ for GF(256). The symbol $^*$ indicates that the value of $S_3^f$ is equal to $S_3^{opt}$.}
	\label{tab:gf_inf_256}
\end{table*}

\begin{table*}[h!]
	\centering
		\begin{tabular}{|r|r r| r r r r | l |} \hline
$d_c$ &	 $S_3^f$	& $S_4^f$	& $M_3$ & $\sigma_3$ & $\Delta_3$ & $R_3$ (\%) & Optimized ($N_g = 1,000$) coefficient's exponents $\{a_i\}_{i=1, \ldots d_c}$ for GF(512)\\ \hline 
3	& $^*$0	  &  15 & 5.02& 4.80  &   1.1 &     0.0 \% & $\{0, 27, 109\}$ \\ 	 
4 & 0 & 100 & 12.4 &  6.5 &  1.9 & 0.0 \% & $\{0, 41, 122, 442\}$ \\  
5 & 0 & 287 & 25.3 &  9.0 &  2.8 & 0.0 \% & $\{0, 39, 155, 320, 436\}$ \\  
6 & 0 & 704 & 45.4 & 11.6 &  3.9 & 0.0 \% & $\{0, 22, 122, 162, 393, 478\}$ \\ 
7 & 3 & 1334 & 74.1 & 14.1 &  5.0 & 4.1 \% & $\{0, 19, 45, 64, 210, 243, 409\}$ \\  
8 & 12 & 2870 & 112.0 & 16.9 &  5.9 & 10.7 \% & $\{0, 19, 75, 119, 159, 228, 312, 367\}$ \\  
9 & 29 & 4576 & 161.0 & 19.5 &  6.8 & 18.0 \% & $\{0, 19, 75, 119, 159, 228, 264, 312, 367\}$ \\ \
10 & 49 & 7599 & 223.7 & 21.7 &  8.0 & 21.9 \% & $\{0, 14, 64, 213, 232, 288, 332, 372, 441, 479\}$ \\  
11 & 77 & 10738 & 299.6 & 24.4 &  9.1 & 25.7 \% & $\{0, 14, 64, 213, 232, 288, 332, 355, 372, 441, 479\}$ \\  
12 & 117 & 15024 & 391.0 & 26.9 & 10.2 & 29.9 \% & $\{0, 13, 27, 77, 226, 245, 301, 345, 366, 386, 454, 492\}$ \\  
13 & 167 & 20232 & 500.2 & 28.0 & 11.9 & 33.4 \% & $\{0, 13, 27, 77, 226, 245, 301, 345, 366, 385, 424, 454, 492\}$ \\  
14 & 233 & 26735 & 627.8 & 31.7 & 12.5 & 37.1 \% & $\{0, 12, 27, 76, 185, 225, 244, 302, 344, 365, 385, 423, 454, 491\}$ \\  
15 & 326 & 34109 & 775.8 & 32.7 & 13.7 & 42.0 \% & $\{0, 12, 27, 63, 76, 185, 225, 244, 302, 344, 365, 385, 423, 454, 491\}$ \\ 
16 & 441 & 43056 & 944.2 & 33.9 & 14.8 & 46.7 \% & $\{0, 12, 27, 63, 76, 185, 225, 244, 302, 332, 344, 365, 385, 423, 454, 491\}$ \\ 
17 & 576 & 53841 & 1135.2 & 36.7 & 15.2 & 50.7 \% & $\{0, 12, 27, 63, 76, 185, 225, 244, 302, 332, 344, 365, 385, 423, 438, 454$ \\ 
     &     &      &     &     &     &     & $ 491\}$ \\   
18 & 733 & 66104 & 1350.8 & 37.6 & 16.4 & 54.3 \% & $\{0, 12, 27, 63, 76, 116, 185, 225, 244, 302, 332, 344, 367, 385, 423, 438$ \\ 
     &     &      &     &     &     &     & $ 454, 491\}$ \\  
19 & 920 & 81171 & 1591.6 & 38.7 & 17.3 & 57.8 \% & $\{0, 10, 25, 41, 78, 98, 110, 125, 161, 174, 214, 283, 323, 342, 400, 430$ \\ 
     &     &      &     &     &     &     & $ 442, 463, 483\}$ \\   
20 & 1130 & 97818 & 1861.7 & 39.9 & 18.3 & 60.7 \% & $\{0, 9, 40, 59, 117, 147, 159, 180, 200, 228, 238, 253, 269, 306, 326, 338$ \\ 
     &     &      &     &     &     &     & $ 353, 389, 402, 443\}$ \\ \hline 
		\end{tabular}
	\caption{List of optimized ($N_g$ = 5000) sets of coefficient's exponents $\{a_i\}_{i=1, \ldots d_c}$ for GF(512). The symbol $^*$ indicates that the value of $S_3^f$ is equal to $S_3^{opt}$.}
	\label{tab:gf512}
\end{table*}

\begin{table*}[h!]
	\centering
		\begin{tabular}{|r|r r| r r r r | l |} \hline
$d_c$ &	 $S_3^f$	& $S_4^f$	&$M_3$ & $\sigma_3$ & $\Delta_3$ & $R_3$ (\%) & GF(1024) \\ \hline
3 & $^*$0 & 3 & 3.6& 4.3  &  0.8 & 0.0 \% & $\{0, 105, 433, 918\}$ \\  
4 & 0 & 57 &  8.5 &  6.2 &  1.4 & 0.0 \% & $\{0, 105, 433, 918\}$ \\  
5 & 0 & 182 & 17.6 &  8.7 &  2.0 & 0.0 \% & $\{0, 30, 328, 358, 448\}$ \\  
6 & 0 & 428 & 31.2 & 11.2 &  2.8 & 0.0 \% & $\{0, 32, 125, 291, 672, 729\}$ \\  
7 & 0 & 1197 & 50.8 & 14.2 &  3.6 & 0.0 \% & $\{0, 67, 208, 592, 685, 829, 956\}$ \\  
8 & 0 & 1680 & 76.6 & 16.7 &  4.6 & 0.0 \% & $\{0, 38, 86, 585, 640, 728, 776, 828\}$ \\  
9 & 3 & 3263 & 110.4 & 19.8 &  5.4 & 2.7 \% & $\{0, 27, 193, 228, 481, 520, 681, 880, 937\}$ \\  
10 & 9 & 5034 & 152.8 & 22.7 &  6.3 & 5.9 \% & $\{0, 35, 121, 288, 327, 489, 554, 744, 830, 888\}$ \\  
11 & 14 & 7681 & 205.4 & 25.8 &  7.4 & 6.8 \% & $\{0, 35, 288, 327, 391, 489, 554, 687, 744, 831, 888\}$ \\  
12 & 24 & 12166 & 268.0 & 28.8 &  8.5 & 9.0 \% & $\{0, 26, 161, 196, 230, 449, 488, 552, 650, 848, 905, 992\}$ \\  
13 & 37 & 16952 & 341.7 & 31.9 &  9.5 & 10.8 \% & $\{0, 24, 219, 258, 322, 420, 618, 675, 762, 793, 819, 954, 989\}$ \\ 
14 & 57 & 22586 & 429.2 & 35.0 & 10.6 & 13.3 \% & $\{0, 24, 219, 258, 322, 420, 576, 618, 676, 762, 793, 819, 955, 990\}$ \\  
15 & 89 & 28936 & 530.0 & 37.8 & 11.7 & 16.8 \% & $\{0, 24, 133, 219, 258, 322, 420, 576, 618, 676, 762, 793, 819, 955, 990\}$ \\ 
16 & 121 & 37290 & 644.9 & 41.3 & 12.7 & 18.8 \% & $\{0, 24, 133, 219, 258, 321, 420, 546, 575, 618, 675, 761, 793, 819, 954, 991\}$ \\  
17 & 173 & 46991 & 775.8 & 44.2 & 13.6 & 22.3 \% & $\{0, 24, 52, 133, 219, 258, 321, 420, 546, 575, 618, 675, 761, 793, 819, 954$ \\ 
     &     &      &     &     &     &     & $ 991\}$ \\ 
18 & 234 & 58191 & 922.2 & 47.1 & 14.6 & 25.4 \% & $\{0, 24, 52, 133, 219, 258, 321, 420, 546, 575, 618, 675, 761, 793, 819, 888$ \\ 
     &     &      &     &     &     &     & $ 954, 991\}$ \\  
19 & 311 & 71377 & 1087.1 & 50.2 & 15.5 & 28.6 \% & $\{0, 24, 52, 133, 219, 258, 321, 420, 518, 546, 575, 618, 675, 761, 793, 819$ \\ 
     &     &      &     &     &     &     & $ 888, 954, 991\}$ \\  
20 & 395 & 88329 & 1270.0 & 52.6 & 16.6 & 31.1 \% & $\{0, 16, 126, 155, 198, 255, 341, 373, 398, 469, 534, 571, 603, 627, 655, 736$ \\ 
     &     &      &     &     &     &     & $ 799, 822, 861, 925\}$ \\ \hline 
\end{tabular}
	\caption{List of optimized ($N_g$ = 5000) sets of coefficient's exponents $\{a_i\}_{i=1, \ldots d_c}$ for GF(1024). The symbol $^*$ indicates that the value of $S_3^f$ is equal to $S_3^{opt}$.}
	\label{tab:gf1024}
\end{table*}

\label{sec:4}

\bibliographystyle{IEEEtran}

\bibliography{protograph}

\end{document}